# Nonlinear signaling on biological networks: the role of stochasticity and spectral clustering

Gonzalo Hernandez-Hernandez[1], Jesse Myers[1], Enric Alvarez-Lacalle[2], Yohannes Shiferaw[1]

[1]Department of Physics & Astronomy, California State University, Northridge

[2]Physics Department, Universitat Politecnica de Catalunya, BarcelonaTech, Barcelona, Spain

**Abstract**

Signal transduction within biological cells is governed by networks of interacting proteins. Communication between these proteins is mediated by signaling molecules which bind to receptors and induce stochastic transitions between different conformational states. Signaling is typically a cooperative process which requires the occurrence of multiple binding events so that reaction rates have a nonlinear dependence on the amount of signaling molecule. It is this nonlinearity that endows biological signaling networks with robust switch like properties which are critical to their biological function. In this study we investigate how the properties of these signaling systems depend on the network architecture. Our main result is that these nonlinear networks exhibit bistability where the network activity can switch between states that correspond to a low and high activity level. We show that this bistable regime emerges at a critical coupling strength that is determined by the spectral structure of the network. In particular, the set of nodes that correspond to large components of the leading eigenvector of the adjacency matrix determines the onset of bistability. Above this transition the eigenvectors of the adjacency matrix determine a hierarchy of clusters, defined by its spectral properties, which are activated sequentially with increasing network activity. We argue further that the onset of bistability occurs either continuously or discontinuously depending upon whether the leading eigenvector is localized or delocalized. Finally, we show that at low network coupling stochastic transitions to the active branch are also driven by the set of nodes that contribute more strongly to the leading eigenvector. However, at high coupling transitions are insensitive to network structure since the network can be activated by stochastic transitions of a few nodes. Thus, this work identifies important features of biological signaling networks that may underlie their biological function.

**Introduction**

A wide variety of intracellular processes are governed by signal transduction between different components of the cell[1, 2]. Typically, signaling occurs at proteins which change their conformational state in response to the presence of a signaling molecule. These molecules are emitted at a source and are transported to the location of the receptor, after which they proceed to trigger a sequence of events leading to a cellular response. Using this mechanism the cell is able to control signals in space and time in order to regulate biological function[3]. Signaling networks which utilize this system are, for example, protein-protein interaction networks, transcriptional regulation, and excitation-contraction coupling in muscle tissue[4-9]. In many of these systems signal transduction is a cooperative process where multiple binding events have to occur in order to trigger a response. Thus, the response of a receptor is typically a nonlinear function of the local concentration of the signaling molecule. Also, receptors are often closely packed into clusters so that a response at one receptor influences the behavior of a nearby receptor. This architecture allows the group of receptors to amplify small concentration changes of the signaling molecule. In this way signal transduction systems can perform robust switch like operations similar to that found in electronic systems[3].

During signaling receptors modulate their transition rates between discrete conformational states. Thus, signaling is an inherently stochastic process which is characterized by probability distributions of the various accessible states. In many cases these probabilities are typically sigmoid functions which relate the response of a receptor to the concentration of the signaling molecules[4]. Another important feature of cellular signaling transduction is that the motion of the signaling molecule can be complex. This is because the intracellular space is filled with organelles and various obstructions so that a diffusing molecule has to navigate between complex intracellular structures. Also, the transmission of signals occurs through a hierarchy of intracellular transport mechanisms. For instance, in many systems the signaling molecule moves as cargo along microtubules, where molecular motors shuttle the signaling molecule between the source and target[10, 11]. In this way signals can be transmitted between distant regions of the cell. Thus, the connections between signaling proteins can be complex and is likely an important component of the biological function.

Given that signal transduction involves a large number of receptors with complex connections it is natural that the tools of network science can be applied to understand their dynamics. To date there has been a great deal of work in this direction and key insights have been gained using this interdisciplinary approach[2, 7, 9]. In this study we explore the general behavior of signaling networks in which nodes are regulated by reaction rates that have a nonlinear dependence on the states of connected nodes. Our main finding is that these networks exhibit bistability where the system can reside either in an active branch, where a large number of nodes are activated, or an inactive branch in which the network is largely quiescent. We analyze the nature of the bistability and show that the onset, and stochastic transitions between the stable branches, are dictated by the community structure of the network. While this work is motivated by cellular signaling networks the main results may also be applied in a wide variety of contexts. The fundamental requirement is simply that nodes on the network communicate in a cooperative fashion so that the local response is a nonlinear function of the activity of connected nodes. Potential applications of this work outside of signaling networks are, for example, in rumor spreading [12] where a member of the population acquires a belief with a probability that increases nonlinearly with the number of people who transmit that belief. This assumption is based on classic experiments in human cognition which have shown that the beliefs of individuals is strongly dependent on the number of people who repeat that belief [13, 14]. This implies that there should a range of contexts where the response of nodes is inherently nonlinear. Thus, the work presented here can potentially be applied to a broad range of complex systems where information spread is both stochastic and nonlinear.

**A network of stochastic signaling units**

In this study we develop a network model that shares basic features with signaling systems observed in biological cells. In particular, our model will be based on the Ca signaling system which is utilized to regulate a vast number of subcellular processes such as muscle contraction and gene transcription[6, 15]. In this model each node corresponds to a biological receptor which responds in a nonlinear fashion to inputs from connected receptors. Specifically, we consider a network of $N$ nodes where node $i$ is described using a state variable $\eta_i$, which can be $\eta_i = 0$ (inactive) or $\eta_i = 1$ (active). Each node undergoes stochastic dynamics with a two state reaction scheme

$$0 \underset{\beta_i}{\overset{\alpha_i}{\rightleftharpoons}} 1 , \tag{1}$$

where $\alpha_i$ is the activation rate, and where $\beta_i$ is the rate of inactivation. The forward rate is given by $\alpha_i = g_i c_i^\gamma$ where $c_i$ is the concentration of the signaling molecule at node $i$, and where $g_i$ is a constant proportional to the excitability of that node. Here, $\gamma$ is an exponent which controls the degree of sensitivity of that node to changes in the concentration of the signaling molecule. In the Ca system receptors typically have multiple binding sites so that activation of receptor proteins is a cooperative process which requires that $\gamma > 1$. This feature is crucial in this study since it endows the system with nonlinear properties. The local signaling concentration is given by

$$c_i(t) = c_o + r \sum_{j \neq i} A_{ij} \eta_j(t) \tag{2}$$

where $c_o$ is a constant that sets the background activity of each site in the absence coupling. Coupling between network nodes is dictated by the matrix $A_{ij}$ which gives a measure of how site $j$ influences the response at site $i$. In our previous study we studied the case where connections depended on the distance between nodes [16]. Here, we will consider networks with long range connections that can be applied to more general signaling networks. For simplicity we consider the case where $A_{ij} = 1$ if there is a connection between node $i$ and $j$, and $A_{ij} = 0$ if there is no connection. In this case $A_{ij}$ corresponds to the network adjacency matrix. Also, the parameter $r$ gives the strength of connection between nodes. We stress here that this network model is general in the sense that it can describe a wide range of systems in which communication between nodes is mediated by the amount of an agent, in this case the concentration $c_i$, which regulates the local response in a nonlinear manner. This model is similar to the susceptible-infected-susceptible (SIS) model that is used to describe disease spreading[17]. However, in the SIS model $c_o = 0$ and $\gamma = 1$ [18, 19], so that a node can only be infected by a connected node, and the response is a linear function of the network activity.

In this study we will simplify the system and treat each node as identical so that $g_i = g$ and $\beta_i = \beta$. Without loss of generality we can chose time units such that $\beta = 1$, so that the forward rate is then simplified to $\alpha_i = \eta \left(1 + \xi \sum_{j \neq i} A_{ij} \eta_j(t) \right)^\gamma$, where $\eta = g c_o^\gamma / \beta$ and $\xi = r/c_o$. Thus, the network dynamics is determined by two dimensionless parameters, $\eta$ which is a measure of the excitability of nodes, and $\xi$ which denotes the strength of coupling. Also, we will focus on the regime where there is low background activity so that $c_o$ is small, and $\eta \ll 1$ and $\xi \gg \eta$. To describe our network of nodes we denote the number of connections to site $i$ as $k_i = \sum_{j \neq i} A_{ij}$. We can also define the average number of connections to all nodes in the network as $\langle k \rangle = (1/N) \sum_{i=1}^{N} k_i$. To compare different networks, it is convenient to study system properties as a function of the variable $s = \xi \langle k \rangle$. This parameter features prominently in our analysis and will be referred to as the coupling strength.

**Dynamics on the Erdos-Renyi network**

As a starting point we will first consider the dynamics of nodes on an Erdos-Renyi (ER) network [20] in which the probability that two nodes are connected is given by a fixed probability $q$. To characterize the activity of the network we compute the fraction of nodes that are active ($\eta_i = 1$) at time $t$ given by

$$p(t) = \frac{1}{N} \sum_{i=1}^{N} \eta_i(t) \tag{3}$$

To explore the range of possible behaviors we will consider an ensemble of initial conditions where the activity level of the network is varied. To do this we chose initial conditions such that $\eta_i = 1$ at site $i$ with probability $h$, and $\eta_i = 0$ otherwise. Here, we simulate $K = 10$ initial conditions where $h = i/K$, and where $i = 0,1,..,K$. In this way we explore the time evolution of a range of random initial conditions from the fully inactive to fully active network. For the ER network we note that there is a percolation transition at $q_c \approx 1/N$, and we will consider networks both above and below the percolation threshold. In Figure 1A-C we plot $p(t)$ vs $t$ for a three different values of the connectivity $s$, and with fixed excitability $\eta = 0.005$. Here, we consider a system of $N = 500$ nodes and with $q = 0.03$ which is above the percolation threshold at $q_c = 0.002$. Our simulations reveal that there are three qualitatively different behaviors. For low $s$ (Figure 1A) we find that all initial conditions evolve towards a steady state where $p(t) \sim 0$ in a time scale roughly $\tau_R \sim 10$. For larger $s$ (Figure 1B) we find that the system is bistable and can evolve, depending upon initial conditions, to both an active ($p(t) \gg 0$) and inactive branch ($p(t) \sim 0$). Finally, for even larger $s$ (Figure 1C) the inactive branch dissapears and the system evolves to the active branch with $p(t) \sim 1$ on a time scale shorter than the relaxation time $\tau_R$. To visualize the possible steady states of the system we compute the quantity $p_\infty = p(T)$, where $T$ is taken to be much longer than the relaxation time $\tau_R$. In Figure 1D we show a density plot of the distribution of $p_\infty$ for the last 50 time units of a simulation up to $T = 100$. Here, the steady state data for all $K = 10$ initial conditions is included in order to give a global picture of the possible steady states of the system. Our results show that for $s < s_a$ all initial conditions evolve to a low activity state where $p(t) \sim 0$. For $s > s_a$ a bistable regime appears where the system evolves to either an active ($p(t) \gg 0$) or inactive ($p(t) \sim 0$) phase depending upon the initial conditions. Finally, for $s > s_b$ we find that all initial conditions evolve to the active branch. Now, below the percolation threshold we find that the network dynamics is substantially different. In Figure 2A-C we plot $p(t)$ vs $t$ for a range of connectivity, and for a network with $q = 0.0015$. To visualize the possible steady states, in Figure 2D, we show a density plot of the distribution of $p_\infty$ vs $s$. In this regime we find that the steady state activity of the network increases gradually above a critical connectivity $s_c$ and no bistability is observed.

An important feature of our signaling network is that the activity level can have distinct steady states. This feature is important since the biological function of these networks is likely dictated by the dynamics towards and between these equilibrium states. Here, we will consider the dynamics of the network and explore how the active state is reached from an initially inactive state where $p(t) = 0$. As a starting point we first consider a network with $q > q_c$ and a range of connectivity $s > s_a$ such that the active branch is stable. In this case we find that if stochastic fluctuations drive $p(t)$ to a threshold $p_{th} = 0.25$ then the system will almost certainly evolve to the active branch. Thus, we can define a waiting time $t_w$ as the time for an initially inactive system, with $p(t) = 0$, to reach the threshold at $p(t_w) = p_{th}$. In Figure 3A we plot the average waiting time $\langle t_w \rangle$ as a function of the connectivity $s$. Here, we find that as the parameter $s$ is decreased from above then the mean waiting time $\langle t_w \rangle$ increases exponentially. This result shows that the stable branch is only observed providing $\langle t_w \rangle \gg T$, and therefore that the critical onset at $s = s_b$ is sensitive to the simulation time. Indeed, as $s$ decreases just below $s_b$ the mean waiting time is substantial ($\langle t_w \rangle \sim 2000$), and decreasing $s$ further leads to prohibitively long simulation times. Thus, for a simulation time of $T \gg \tau_R$ the system exhibits bistability over a broad range of $s$. On the other hand, below the percolation threshold there is no gap between the active and inactive phase. In this case we can compute the mean waiting time to reach an arbitrarily chosen activity level $p_{eq}$, providing that $p_{eq} < p_\infty$ In our simulation, shown in Figure 3B, we pick $p_{eq} = 0.2$, and show that $\langle t_w \rangle$ increases as $s$ is decreased to the level where $p_{th} \approx p_\infty$. Thus, below the percolation threshold spreading activity increases gradually to a well defined steady state and threshold behavior is not observed.

To summarize, an ER network above the percolation transition exhibits bistability where initial conditions evolve towards an active or inactive steady state in a time scale $\tau_R \sim 10 - 20$. In this regime an inactive network can undergo stochastic transitions to the active state on time scales $\langle t_w \rangle \gg \tau_R$, and the exponential growth of this time scale as $s \rightarrow s_b$ ensures that this bistability is observed for long simulation times $T$. Thus, above the percolation transition the ER network behaves like a perfect switch with extremely rare transitions between distinct states. On the other hand, below the percolation transition the network response is graded and does not exhibit bistability. In this case long dwell times between distinct network states is not observed, and the system evolves to a final state that is insensitive to initial conditions.

**Dynamics on the Barabasi-Albert network**

In this study we will also explore dynamics on a Barabasi-Albert (BA) network [21]. These networks have a scale free structure which has been observed in a variety of biological signaling networks [8]. Briefly, a BA network is formed by adding a sequence of $m$ new nodes to an existing network of nodes. Each new node added is then connected preferentially to the nodes with larger degree. In Figure 4A-E we plot trajectories of the activity $p(t)$ for a BA network with $m = 5$, and in Figure 4F we plot the distribution of the steady state $p_\infty$. Here, we find that all initial conditions evolve to the inactive state providing our connectivity is below a critical value $s = s_a \sim 40$ (Figure 4A). For $s_a < s < s_b$ we find that the system is bistable and stochastic transitions between the active and inactive branch are observed. Here, transitions between these states occur over time scales substantially longer than the relaxation time of the network. Finally, for $s > s_b$ all initial conditions evolve to the active branch (Figure 4D). Finally, we note that at onset the active branch emerges at a gap defined as $\Delta = p_\infty(s_a)$ (Figure 4F). This gap is an important characteristic of the BA network and will feature prominently in our subsequent analysis.

To explore dynamics on the BA network we have computed the mean life time of the active and inactive network states. As a starting point, we simulate the evolution of an initially inactive state which is reached with initial conditions $\eta_i = 0$. In Figure 5A (blue line) we plot the mean waiting time $\langle t_w \rangle$ for an inactive system to transition to the active branch as a function of the connectivity $s$. As in the ER network the waiting time $\langle t_w \rangle$ increases exponentially as $s$ is decreased into the bistable regime. However, unlike the ER network, transitions to the active branch occur on shorter time scales which can be observed deep into the bistable regime. We have also computed the average waiting time for an active network to transition to the inactive state. In this case (Figure 5A, black line) the average waiting time $\langle t_a \rangle$ increases exponentially as $s$ is increased above $s_a$. Thus, the network is bistable with stochastic transitions between two distinct network states. Now, for different network connectivity $m$ we find qualitatively similar results. In Figure 5B we plot the gap at onset $\Delta$ vs $m$, showing that the two stable branches gradually merge as the network connectivity $m$ is decreased. In Figure 5C we plot $p_\infty$ vs $s$ for $m = 2$ showing that the onset of the active branch at $s = s_a$ emerges gradually although a small region of bistability is observed. In this case stochastic fluctuations on the network smear the gap and the transition appears continuous.

**Deterministic mean field theory**

In order to understand the basic features of the network dynamics we have developed a simplified mean field theory of the system. Following our previous work, we define $p_i(t)$ to be the probability that the $i^{th}$ node is active ($\eta_i(t) = 1$) at time $t$. The stochastic dynamics is then governed by the master equation

$$p_i(t + \tau) = (1 - p_i(t))(\alpha_i \tau) + p_i(t)(1 - \beta \tau) \tag{4}$$

where $\alpha_i \tau$ and $\beta \tau$ are the transition probability that node $i$ makes a $0 \to 1$ or $1 \to 0$ transition in the time interval $\tau$. To proceed we make the approximation that state variables can be replaced by their averages so that $\eta_j \approx p_j$. In the continuum limit the master equation then simplifies to the system of equations

$$\frac{dp_i}{dt} = \eta \left(1 + \xi \sum_{j \neq i} A_{ij} p_j \right)^\gamma (1 - p_i) - p_i, \tag{5}$$

which gives a set of $N$ coupled nonlinear equations for the site probability $p_i$. Here, we note that in the limit of low background signaling $c_o \to 0$, in which $\xi \gg 1$, and for the case $\gamma = 1$, our deterministic equations reduce to the deterministic limit of the SIS model. In this study we will focus mainly on the nonlinear regime $\gamma > 1$ which is relevant to signaling in biological networks.

As a starting point, we will first consider the spatially homogeneous case where $p_i = p$ and $\sum_{j \neq i} A_{ij} = \langle k \rangle$. The mean-field equations then reduce to a single equation

$$\frac{dp}{dt} = \eta(1 + sp)^\gamma (1 - p) - p, \tag{6}$$

where $s = \xi \langle k \rangle$. This equation possesses stationary points that satisfy the algebraic condition

$$\eta(1 + sp)^\gamma (1 - p) - p = 0. \tag{7}$$

In Figure 6A we plot the solution to Eq. (7), with $\eta = 0.005$ and $\gamma = 2$, showing that the system is monostable for $s < s_1$, bistable in the range $s_1 < s < s_2$, and monostable for $s > s_2$. To estimate the turning points we note that $\eta \ll 1$, so that in this regime the transitions between the monostable and bistable states can be well approximated by $s_1 \approx 2/\sqrt{\eta}$ and $s_2 \approx 1/4\eta$.

To study the full mean field equations we will solve Eq. (5) on both the ER and BA networks. To keep track of the activity of the network we compute the quantity

$$p(t) = \frac{1}{N} \sum_{i=1}^{N} p_i(t) \ , \tag{8}$$

which corresponds to the activity level of the network. To characterize the steady state we define $p_\infty = p(T)$ where $T$ is taken to be much longer than the relaxation time of the system. In Figure 6B we plot $p_\infty$ as a function of $s$ (red circles) for the ER network above the percolation threshold. On the same graph we also plot the solution of the mean field equations shown in Figure 6A (blue dashed line). Indeed, we find that the steady state activity level $p_\infty$ is well approximated by the stable branches of the steady state solution of the mean field approximation. On the same graph we have also plotted the steady state probability $p_\infty$ for an ER network below the percolation transition ($q = 0.0015$). In this case we observe markedly different behavior. Here, we find that for $s < s_3$ the steady state evolves to an inactive state with $p_\infty \sim 0$ independently of initial conditions. For $s > s_3$ the network evolves to distinct steady states which depend on the initial conditions picked. However, this regime of multistability ends abruptly at a critical connectivity and the system again becomes monostable.

We have also solved the mean field equations on the BA network. In Figure 6C we plot $p_\infty$ vs $s$ for a network with connectivity $m = 5$. In this case the mean field equations exhibit bistability with the onset of the active branch at a critical $s = s_1$, where there is a well defined gap $\Delta$ between the two branches. Furthermore, both branches remain stable up to $s = s_2$ after which the inactive branch is no longer a steady state solution. Now, for different network connectivity $m$ we find similar results but with a gap $\Delta$ that grows approximately linearly with $m$ (Figure 6D). Here, we point out that the mean field equations on a BA network only displays bistability and never multistability. To check this observation we have repeated the simulation up to $N = 5000$ nodes, and find that indeed only bistability is observed with features similar to that shown in Figure 6C. Also, this bistability appears to be robust even in the case where the activation and inactivation rates are heterogeneous. Thus, the power law structure of the BA network ensures a bistable response over a wide range of conditions.

**Onset of the active branch is dictated by the spectral structure of the network**

In this section we analyze our mean field equations in order to develop a phenomenological theory of the network activity. Firstly, we note that our mean field theory, along with the assumption that all nodes are equivalent, served as a good approximation to the steady state dynamics of the ER network above the percolation transition. This result is expected since for $q > q_c$ an ER network is dominated by a giant component where all nodes share the same average connectivity, and can thus be treated as equivalent. Now, in the case where $q < q_c$ this assumption will not apply. In this case it is reasonable to approximate the network as a collection of clusters, where nodes within each cluster have a higher than average connectivity, but nodes in different clusters are unconnected or weakly connected. Within each cluster we can apply our mean field approximation so that the steady state of the $l^{th}$ cluster is governed by the algebraic condition

$$\eta(1 + \xi \langle k \rangle_l p_l)^\gamma (1 - p_l) - p_l = 0 \tag{9}$$

where $\langle k \rangle_l$ denotes the average connectivity of the $l^{th}$ cluster, and where $p_l$ denotes the fraction of active nodes of that cluster. The onset of bistability of the $l^{th}$ cluster then occurs at $\xi \langle k \rangle_l \approx 2/\sqrt{\eta}$ and transitions back to monostability when $\xi \langle k \rangle_l \approx 1/4\eta$. Now, note that the full network connectivity is defined by the parameter $s = \xi \langle k \rangle$ where $\langle k \rangle$ is the average connectivity of all $N$ nodes. Thus, the $l^{th}$ cluster is bistable for $s_1^l < s < s_2^l$ where

$$s_1^l \approx \frac{2}{\sqrt{\eta}} \left( \frac{\langle k \rangle}{\langle k \rangle_l} \right) \tag{10}$$

and

$$s_2^l \approx \frac{1}{4\eta} \left( \frac{\langle k \rangle}{\langle k \rangle_l} \right) . \tag{11}$$

To proceed, let us now order clusters within the network such that $\langle k \rangle_1 > \langle k \rangle_2 > \cdots > \langle k \rangle_K$ where $K$ is the number of distinct clusters. Here, the cluster $l = 1$ has the largest average connectivity and will be referred to as the leading cluster. Now, since $s_1^1 < s_1^2 < \cdots < s_1^K$ then the leading cluster will be the first to bifurcate to an active branch at the onset $s_3 = s_1^1$. At this connectivity, the network, depending upon initial conditions, can evolve so that the leading cluster resides on the active branch at steady state. Now, if we denote the number of nodes in the leading cluster to be $n_1$, then the active branch will have the steady state probability $p_\infty(s_a^1) \approx n_1/N$. Therefore, the gap between the active and inactive branch is related to the size of the leading cluster according to

$$\Delta \approx \frac{n_1}{N} . \tag{12}$$

Thus, the gap gives a direct estimate of the fraction of nodes that belong to the leading cluster of the network. Now, as the connectivity is increased further then more clusters will develop an active branch so that multiple clusters will be bistable. Hence, if $s > s_1^i$ for $i < 1, 2, \ldots, M$ then all these clusters will contribute to the active branch and the number of distinct steady states is $2^M$. This result explains the multistability observed in Figure (6B), where different steady states are reached depending upon initial conditions.

**Onset of active branch dictated by leading eigenvectors of the network adjacency matrix**

Our phenomenological theory predicts that the onset of the active branch is determined by the leading cluster in the network. To identify this cluster we will apply standard results in spectral clustering which identify clusters of nodes from the eigenvectors of the network adjacency matrix[22, 23]. In particular, we will identify the leading cluster as the set of nodes which correspond to the large components of the eigenvector $\vec{\phi}_1$ corresponding to the largest eigenvalue of $A_{ij}$. In Figure 7A (red circles) we plot the ordered components of the leading eigenvector ($\phi_{1i}$) for a BA network with $m = 2$. Here, components are normalized so that the maximum component is 1. In this case we observe that there are roughly 10 nodes with components substantially larger than the typical component of the eigenvector, and it is reasonable to identify the leading cluster as the set of nodes such that $\phi_{1i} > 0.2$. To compare these nodes with the network activity we compute the running average at each node defined as

$$\bar{\eta}_i = \frac{1}{T - T_i} \sum_{t=T_i}^{T} \eta_i(t) \, \Delta t \tag{13}$$

where $T_i$ is time at which we begin the time average, and where $T$ is the final time. In these simulations we wait for times much longer than the relaxation time of the system $\tau_R \sim 20$, so that the system reaches steady state before we initiate the time average. Thus, $\bar{\eta}_i$ gives a measure of the activity level at node $i$ at steady state. In Figure 7A (blue squares) we show $\bar{\eta}_i$ vs $i$ for a network connectivity of $s = 40$ which is close to the onset at $s_a$ (see Figure 5C). Here, we have again ordered the components $\bar{\eta}_i$ and normalized the vector so that the largest component is one. To compare activity levels with the leading eigenvector, in Figure 7B we map the network of 500 nodes. Here, we identify all nodes with $\phi_{1i} > 0.2$ and place them within a circle at the center of the network map. On the same arrangement we designate in red the high activity sites with $\bar{\eta}_i > 0.5$. Indeed, we find that the most active parts of the network coincide with the leading cluster which can be identified using the leading eigenvector. To quantify the relationship between eigenvector and steady state we have also computed the Pearson correlation defined as

$$C(\vec{\phi}_1, \vec{\eta}) = \frac{\sum_i^N (\bar{\eta}_i - \langle\bar{\eta}\rangle)(\phi_{1i} - \langle\phi_1\rangle)}{\sqrt{\sum_i^N (\bar{\eta}_i - \langle\bar{\eta}\rangle)^2} \sqrt{\sum_i^N (\phi_{1i} - \langle\phi_1\rangle)^2}}, \tag{14}$$

which gives a measure of the correlation between activity level and eigenvector. Here, $C \sim 0$ implies no correlation while $C \sim 1$ corresponds to high correlation. Computing this quantity for our network yields $C = 0.73$ which shows that the active branch at onset is highly correlated with the leading eigenvector of the system. This result confirms our phenomenological theory which predicted that the onset of the active branch is determined by the leading cluster of the network.

In the ER network we find similar results. For this network we analyze the case where $q < q_c$ in which we have a continuous transition to the active branch at $s_c \approx 11$. In Figure 8A we plot the steady state activity $\bar{\eta}_i$ aftering ordering and on the same graph show the components of the leading eigenvector. Here, we find that most components of $\vec{\phi}_1$ are zero, and that the non-zero components identify isolated clusters of the network which have higher connectivity than average. Thus, empirically we can identify the leading cluster as the set of nodes with $\phi_{1i} > 0.01$. In Figure 8B we plot the nodes where $\bar{\eta}_i > 0.5$, and simultaneously the nodes at which $\phi_{1i} > 0.01$. Indeed, we find that the steady state active sites are correlated with the cluster associated with the leading eigenvector. However, in this case we find that there are also contributions from clusters associated with the eigenvectors that correspond to the 2$^{nd}$ and 3$^{rd}$ largest eigenvalues. This finding is not surprising since the spacing between eigenvalues of the ER network is small for $q <$ $q_c$, it is expected that clusters corresponding to each eigenvector have similar strength making them contribute to the onset of the active branch.

**The structure of the leading eigenvector: localization and delocalization**

A basic prediction of our phenomenological theory is that the onset of the active branch is determined by the clusters associated with the leading eigenvectors of the network adjacency matrix. Therefore, at onset we expect that the structure of the leading eigenvector will determine the transition to the active branch. In particular, if the size of the leading cluster is $n_1$ then the gap between the inactive and active branch at onset is just $\Delta = n_1/N$. Therefore, we can estimate $\Delta$ by counting the components of the leading eigenvector with magnitudes much larger than the typical component. In the case of the ER network $n_1$ is just the number of nodes with components $\phi_{1i} > 0.01$. This simple criterion is remarkably effective to identify groups of highly connected nodes in our ER network which are disconnected from the rest of the nodes (Figure 8B). In Figure 9A we plot $\Gamma = n_1/N$ as a function of the linking probability $q$. This result indicates that $\Gamma$ exhibits a sigmoid dependence on $q$ with a transition point close to the percolation threshold at $q_c \sim 1/N = 0.2 \times 10^{-2}$. Thus, for $q < q_c$ the leading cluster is a small fraction of the network, while for $q > q_c$ it grows to system size due to the emergence of the giant component. In Figure 9B we show the ordered components of the leading eigenvector above and below the percolation threshold. Indeed, we find that for $q < q_c$ the leading eigenvector is localized at a small number of nodes, while for $q > q_c$ the vector is delocalized across all nodes in the system. Therefore, for $q > q_c$ the gap $\Delta$ to the active branch occurs due to a discontinuous jump since the leading cluster associated is effectively system size. However, for $q < q_c$ we have that $n_1 \sim 10$ and $\Delta \sim 0.02$, which is substantially smaller. In fact, in Figure 2D we find that stochastic fluctuations smear this gap so that it appears continuous in our simulations of the steady state.

Now, an important question to address is whether the onset of the active branch is a continuous or discontinuous transition in the infinite system size limit $N \to \infty$. To address this question we have computed $\Gamma$ vs $N$ for an ER network above and below the percolation threshold. In the ER network it is well known that $q_c \approx 1/N$ so that we can remain below the percolation threshold by picking $q = 1/2N$ and above with $q = 3/N$. In Figure 9C & D we plot $\Gamma$ vs $N$ above and below the percolation threshold for system sizes in the range $500 < N < 2000$. Indeed, we find that below the percolation transition the size of the leading eigenvector decreases with increasing system size. However, above the percolation transition the size is comparable to the system size $\Gamma \sim 0.9$ and does not change substantially with increasing system size. These results suggest that in the limit $N \to \infty$ we expect a continuous transition to the active branch providing $q < q_c$, and a discontinous transition for $q > q_c$. Thus, the qualitative nature of the transition is dependent on the localization properties of the leading eigenvector of the network adjacency matrix.

In the case of the BA network we find that the size of the leading eigenvector depends crucially on the connectivity $m$. In Figure 10A we plot $\Gamma$ vs $m$ showing that the size of the leading cluster increases linearly with connectivity $m$. In Figure 10B we plot $\Gamma$ vs system size $N$ at a fixed $m = 5$. Here, we observe that $\Gamma$ decreases with increasing $N$, which is consistent with previous studies, showing that the leading eigenvector of a BA network is localized[24]. This result implies that in the limit $N \to \infty$ then $\Delta \to 0$, so that the transition to the active branch will be continuous. Here, we note that the leading cluster of the BA network is substantially larger than that of the ER network. Thus, for a system of size $N = 500$ a clear gap is observed in the BA network for $m > 2$, while for $q < q_c$ this gap is smeared out by stochastic fluctuations of the network.

**The role of noise**

Our deterministic mean field theory explains several important features of the nonlinear stochastic network. In particular, in the case of the ER network the mean field equations correctly describe the system above the percolation threshold. However, for $q < q_c$ the stochastic model (Figure 2D) does not exhibit the multistability predicted by the mean field equations. To explain this discrepancy, we note that our mean field equations neglect the effect of noise. Indeed, when we include a noise term in our mean field equations we find that the qualitative results of the full stochastic model are reproduced. To understand this result let us consider a network in which there are $M$ disconnected clusters which possess a stable active and inactive branch i.e. these clusters are in the bistable regime. It is then possible to pick initial conditions so that only a fraction of these $M$ clusters will evolve towards the active state. Thus, the deterministic equations can evolve to the possible $2^M$ possible stable states of the network. Now, in the presence of noise then all of these clusters can be activated by stochastic fluctuations. Thus, the steady state active branch we observe in the network dynamics corresponds the case where all $M$ clusters are activated. In this way, the stochasticity of the system reduces the dispersion of the network activity by driving the system to the maximally activated state. However, the role of noise is different in the BA network. Here, we observe robust bistability and noise only induces stochastic transitions between the active and inactive branch. In this case noise only modifies the range of the bistable regime and does not qualitatively change the dependence on the connectivity of the network.

**Spreading dynamics is governed by fluctuations at the leading cluster**

In this section we will briefly address the question of how signals spread on the network of nodes. In effect, we ask how a network that is initially inactive, with $\eta_i = 0$ for all $i$, evolves towards the active branch at steady state. To address this question we will first consider a BA network with $m = 2$ which exhibits the richest dynamical behavior near threshold. In Figure 11A we plot the number of active nodes on the network $n(t)$ as a function of time during a time interval where there is a transition to the active state. Here, we fix $s = 39.8$ which is just above the threshold at $s_a = 37.8$ (see Figure 5C). In this case we observe that the transition is initiated by stochastic fluctuations which can nucleate a cascade of activation leading to the active branch. By studying several initiation events we find that in this case roughly 20 nodes are active before a nucleation event. Thus, transitions between the inactive and active branches are nucleated by the activation of a critical number of nodes $n_c \sim 20$. To uncover the structure of these nodes we identify the active nodes when the system reaches a fixed threshold of $n_c = 20$. For each node we proceed to compute the ratio $P_i = k_i/M$ where $k_i$ is the number of times node $i$ was active when the threshold condition was reached in $M = 2000$ independent simulations. Thus, the vector $\vec{P}$ gives the distribution of active sites at threshold. In Figure 11B we show a map of our system and circle those sites that belong to the leading cluster of the network. On the same graph we highlight in red the 10 nodes at which $P_i$ is largest. Here, we find that near threshold there is a strong correlation between active sites and the leading cluster. Given that the connectivity of the node is strongly correlated with its component in the leading eigenvector, this result indicates that the nodes involved in the nucleation of cascades in the network correlate well with the most connected nodes. However, for large system connectivity $s = 159.2$ we find that this stong correlation is lost. In this regime we see that only one 1-2 nodes need to fire to initiate spreading on the network (Figure 11C). Indeed, a map of active sites at a threshold $n_c = 1$ reveal that nucleation can occur with equal likelihood at any node on the network. In this case stochastic fluctuations induce a full network activation and the nucleation process is independent of the underlying network. To quantify these observations in Figure 12 we compute the Pearson correlation between the leading eigenvector $\vec{\phi}_1$ and the density of active sites at threshold $\vec{P}(t)$. Indeed, we find that the correlation between these quantities decreases monotonically as $s$ is increased from the onset of the active branch. Thus, for low connectivity the leading cluster dictates the sites where the initiation of spreading occurs, while for high connectivity spreading initiation is largely independent of the network structure.

To explain our numerical findings and to stablish a direct link with the leading cluster, and indirectly with the nodes with higher connectivity, we can apply our mean field approximation and study the linear stability of the stable branch. In this case we expand Eq. (5) near the steady state solution $p_i^*$ so that $p_i = p_i^* + u_i$, and where $u_i$ is a small perturbation. Expanding to linear order in $u_i$ gives the linear system of equations

$$\frac{du_i}{dt} = -(1 + \eta (1 + \xi q_i)^2) u_i + 2\eta \xi (1 - p_i^*)(1 + \xi q_i) v_i, \tag{15}$$

where $q_i = \sum_j A_{ij} p_j^*$, and where $v_i = \sum_j A_{ij} u_j$. Thus, the stability of the low activity state is dictated by the system of equations $\dot{u}_i = \sum_j Q_{ij} u_j$ where

$$Q_{ij} = -(1 + \eta(1 + q_i)^2)\delta_{ij} + 2\eta\xi(1 - p_i^*)(1 + \xi q_i) A_{ij}. \tag{16}$$

The time evolution of the system is then given by the linear combination

$$\vec{u}(t) = \sum_k a_k e^{\Gamma_k t} \vec{\psi}_k \quad, \tag{17}$$

where $\Gamma_k$ and $\vec{\psi}_k$ are the eigenvalues and eigenvectors of the matrix $Q_{ij}$. Thus, the quiescent state is unstable when the maximum eigenvalue, which we denote here as $\Gamma_1$, satisfies the condition $\Gamma_1 \geq 0$. To simplify the analysis further we note that the steady state occupation probabilities are small so that $p_i^* \sim 0$, so that $1 - p_i^* \sim 1$. Finally, we will make the assumption that the terms $q_i = \sum_j A_{ij} p_j^*$ do not vary substantially from site to site. Hence, we can replace $q_i \to \langle q \rangle$, where $\langle q \rangle = (1/N) \sum_i q_i$ is the average over the network. This gives an approximation

$$Q_{ij} \approx -(1 + \eta(1 + \xi\langle q \rangle)^2)\delta_{ij} + 2\eta\xi(1 + \xi\langle q \rangle)A_{ij}. \tag{18}$$

The leading eigenvalue can then be approximated by $\Gamma_1 = -(1 + \eta(1 + \xi\langle q \rangle)^2) + 2\eta\xi(1 + \xi\langle q \rangle)\lambda_1$, where $\lambda_1$ is the leading eigenvalue of the network adjacency matrix.

Our linear stability analysis shows that the inactive branch is unstable when $\Gamma_1 > 0$, and that the growth rate of network activity will be well approximated by the leading eigenvector $\vec{\phi}_1$ of the adjacency matrix $A_{ij}$. However, our numerical results show that fluctuations to the active branch also occur in the case $\Gamma_1 < 0$. In this case, while the inactive state is stable, stochastic fluctuations can induce the system to transition between the inactive and active branch. Our key finding here, which was pointed out in our previous study [16], is that these fluctuations occur at the leading cluster. The reason for this is simply that in the stable regime fluctuations are suppressed with a weight factor $\propto \exp(\Gamma_1 t)$. Therefore, it is precisely fluctuations at the leading cluster which decay at the slowest rate and therefore have the greatest likelihood of inducing a stochastic transition between inactive and active states. In fact, a key finding of Figure 12, is that the correlation between the sites of nucleation and the leading eigenvector increases monotonically as the connectivity is decreased into the bistable regime. Thus, while transitions become rarer at low connectivity they also occur with higher likelihood at the leading cluster of the network.

**Discussion**

In this paper we have explored the dynamics of a network where signaling between nodes is stochastic and nonlinear. The important feature that we have identified is that, because of the nonlinear interactions between connected nodes, the network exhibits bistability. This bistability corresponds to an inactive branch where most of the nodes are in the off state, and an active branch where a fraction of the system is active. Our analysis revealed that the onset of the active branch is dictated by the cluster of nodes that correspond to the leading eigenvector of the network adjacency matrix. For the ER network we find that this result implies that the network dynamics above and below the percolation transition is dramatically different. Effectively, below the percolation threshold the leading eigenvector is localized on a few network sites, while above, it is spread throughout the network. Thus, below the percolation threshold the active branch emerges in a continuous fashion, and network activity grows gradually as a function of increasing excitability. On the other hand, above the percolation transition there is a substantial gap between the two branches and the onset of the active branch is discontinuous. Consequently, stochastic transitions between the two stable states of the network are extremely rare and the steady state dynamics of the system is determined mostly by initial conditions. Now, in the case of the BA network we observe similar yet distinct behavior. In this case the leading eigenvector is also localized although the size of the leading cluster is substantially larger than in the case of the ER network. Thus, a finite gap between the active and inactive branches can be observed at least for the system sizes considered in this study. Furthermore, we found that stochastic transitions between the active and inactive branch was readily observable, with dwell times that increased exponential with the network connectivity. Indeed, these stochastic fluctuations occur at nodes that can be identified from the leading eigenvectors of the adjacency matrix.

In the context of biological signaling networks our results have several important implications. The most important is simply that the network connectivity plays a fundamental role in the global behavior of the system. For example, an ER network above the percolation transition essentially behaves like a perfect switch with extremely rare transitions between the active and inactive state. On the other hand, below the percolation transition, the network response to system parameters is continuous and graded. From a biological perspective this result implies that the wiring architecture of the signaling network dictates the system response, and that this behavior should be robust to changes in the properties of individual nodes. Now, in the case of the BA network we find that the system can also exhibit switch like behavior between distinct macroscopic states. A crucial finding is that both the onset and the nucleation dynamics of activation in these networks are dictated by the leading cluster of the network. In the biological context this implies that networks with a BA architecture are particularly susceptible to stochastic nucleation events centered at the leading cluster of the network. In effect, the overall activity level of the network will be driven by certain "key players" which have a higher than average network connectivity. Thus, in these systems it is crucial to identify these key players in order to understand the function of these networks.

In this study we have also investigated the nature of spreading in these nonlinear networks. In particular, we explored how a network that is initially inactive makes the transition to the active branch. Our main finding here is that at high connectivity spreading is initiated, with equal likelihood, at any site in the network. This is because only a few connected nodes need to fire simultaneously in order to induce a large scale transition to the active phase. However, as the connectivity is decreased then the network structure dictates the sites of activation. In this case we find that activation is a nucleation process on the network which tends to occur at the leading cluster. This result is consistent with known results on disease spreading on networks, where it is known that the threshold for spreading is dictated by the leading eigenvalue of the network adjacency matrix[17]. However, in the scenario considered here, the system is bistable so that stochastic transitions can occur to the active phase even in the case when the low activity branch is stable. In this case we find that the leading cluster is even more predictive of the sites of nucleation. This is because the leading cluster is precisely that set of well-connected nodes where random fluctuations are suppressed by the least amount. Therefore, as the network connectivity is reduced into the bistable regime, nucleation of spreading activity becomes exponentially rarer but occurs with increasing likelihood at the leading cluster.

It will be interesting to explore if our findings can be applied to other spreading processes. For example to infectious disease or virus spreading on networks, or to the spread of signals in neuronal activity in in-vitro cultures[25]. Here, the key requirement is that the activation of nodes must be a cooperative process that is a nonlinear function of neighboring states. We point out that in biological signaling networks this requirement is natural since proteins are typically regulated by multiple binding sites. On the other hand, in disease or virus spreading it is generally believed that infection is linearly related to the number of infected sites. However, there may be contagious processes which have a nonlinear dependence on infected nodes. A scenario where the response of a node is probably nonlinear is the study of rumor spreading in a population. In fact, it is well known that an individual's belief is strongly dependent on the number of different sources who repeat that believe as truth. This phenomenon, which is known as the illusion-of-truth[14], will likely introduce a degree of nonlinearity in the interaction between members of a population. In this case, our main results such as bistability, stochastic nucleation, and importance of the leading cluster, will apply, and should explain the main features of the spreading process.

**Acknowledgement**

This work was supported by the National Heart, Lung, and Blood Institute grant RO1HL101196. We thank Dr. Romualdo Pastor-Satorras for fruitful discussions.

**Captions**

**Figure 1.** The network activity $p(t)$ as a function of time for a range of initial conditions. Here, we use an ER network of size $N = 500$ that is above the percolation threshold ($q = 0.03$), and with fixed excitability $\eta = 0.005$. (A) Inactive phase with $s = 24$. (B) Bistable phase with $s = 27$. (C) Active phase with $s = 60$. (D) Density plot of the distribution of steady state probability $p_\infty$ for the last 50 time units of a simulation up to $T = 100$.

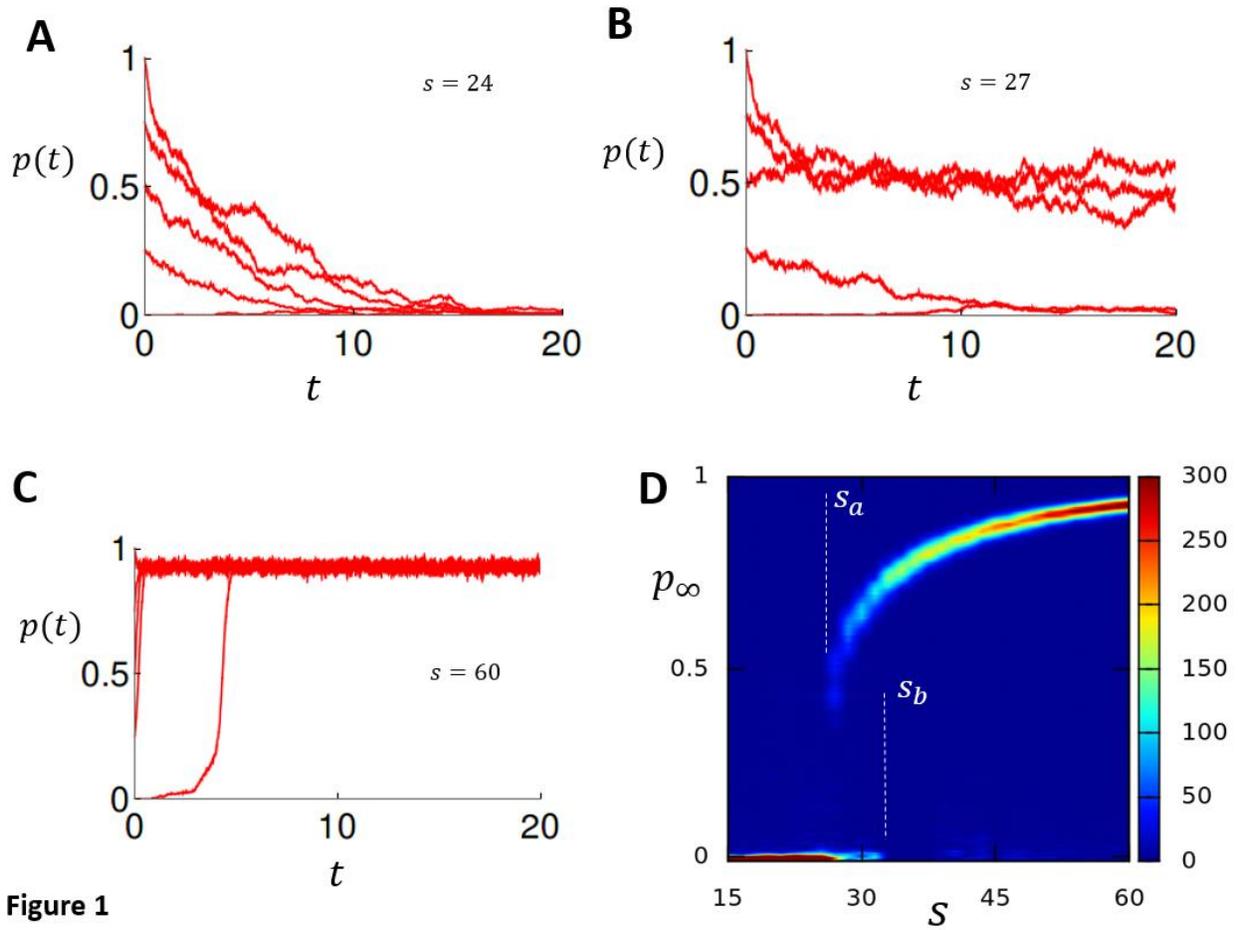

**Figure 1**

**Figure 2.** The network activity $p(t)$ for an ER network below the percolation transition ($q = 0.015$). Network size is $N = 500$ and excitability is fixed at $\eta = 0.005$. (A) Inactive phase with $s = 5.6$. (B-C) Active phase with $s = 15.3$ and $s = 501$. (D) Density plot of the distribution of steady state probability $p_\infty$ for the last 50 time units of a simulation up to $T = 100$.

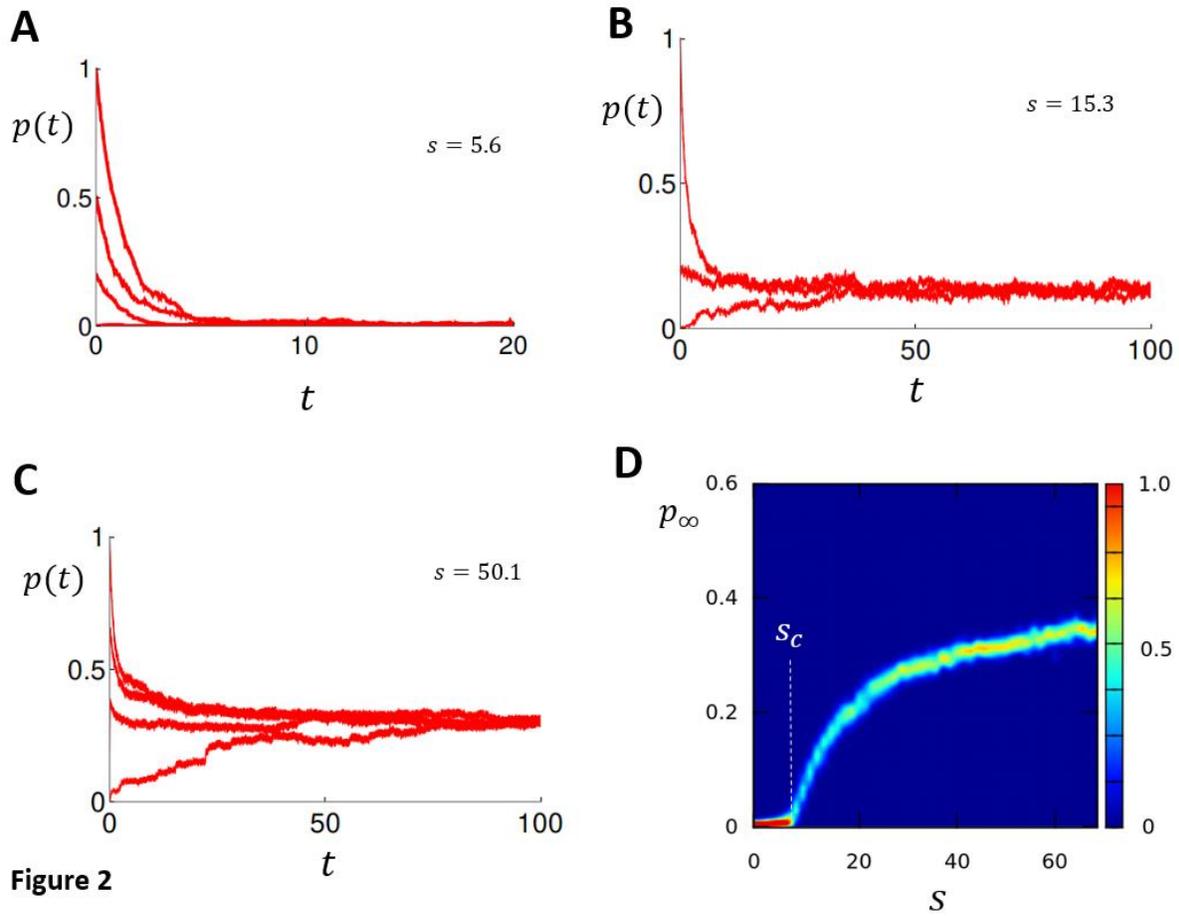

Figure 2

**Figure 3.** Mean waiting time $\langle t_w \rangle$ as a function of the connectivity s. This is the time an inactive system starting with $p(t)=0$ will take to reach the threshold $p(t_w) = p_{th}$. The waiting time is averaged over 1000 independent simulation runs. (A) Average waiting time $\langle t_w \rangle$ for an ER network above the percolation threshold and a threshold $p_{th} = 0.25$. (B) Average waiting time $\langle t_w \rangle$ for an ER network below he percolation threshold and a threshold $p_{th} = 0.20$.

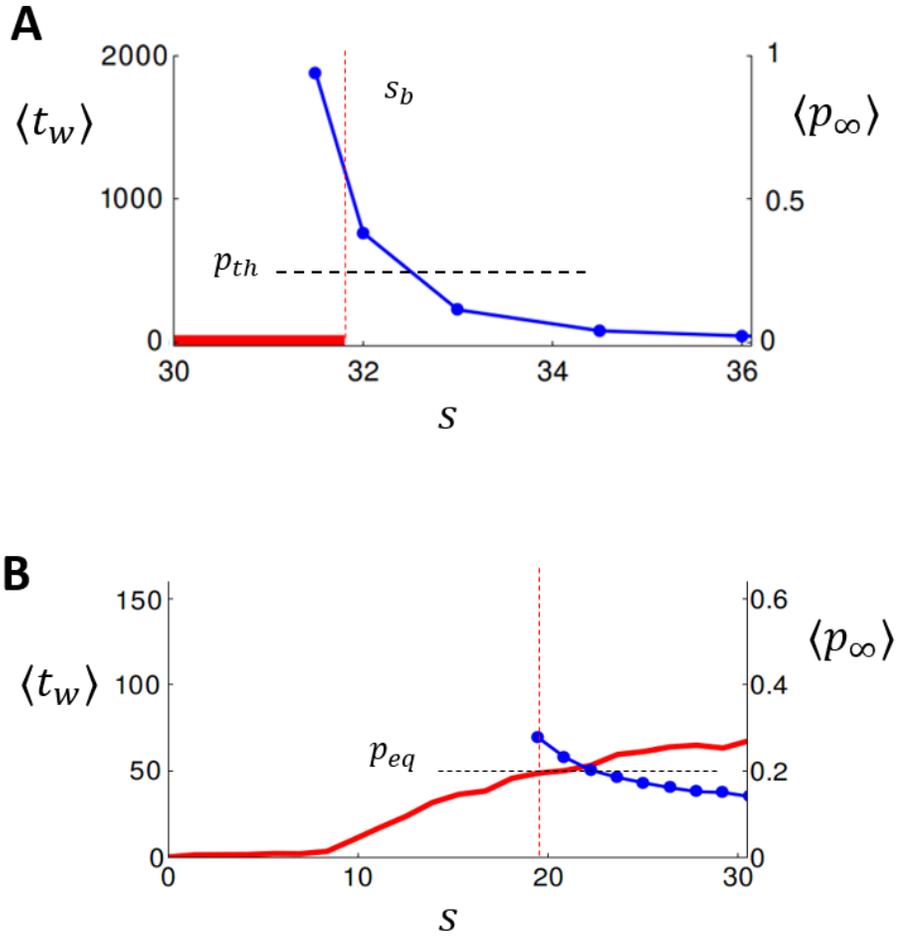

Figure 3

**Figure 4.** The network activity $p(t)$ vs time for a range of initial conditions. Here, we simulate a BA network of size $N = 500$, with connectivity $m = 5$, and fixed excitability $\eta = 0.001$. (A) Inactive phase with $s = 20$. (B-D) Bistable regime with $s = 43.2, 44.2$, and $45.7$, showing stochastic transitions from the inactive to active phase. Note the long dwell times in each state. (E) Time evolutions of system in the active regime ($s = 99.4$), where all initial conditions evolve to a high activity state. (F) Density plot of the distribution of steady state probability $p_\infty$ for the last 100 time units of a simulation up to $T = 200$. The gap separating the inactive and active branch is denoted as $\Delta$.

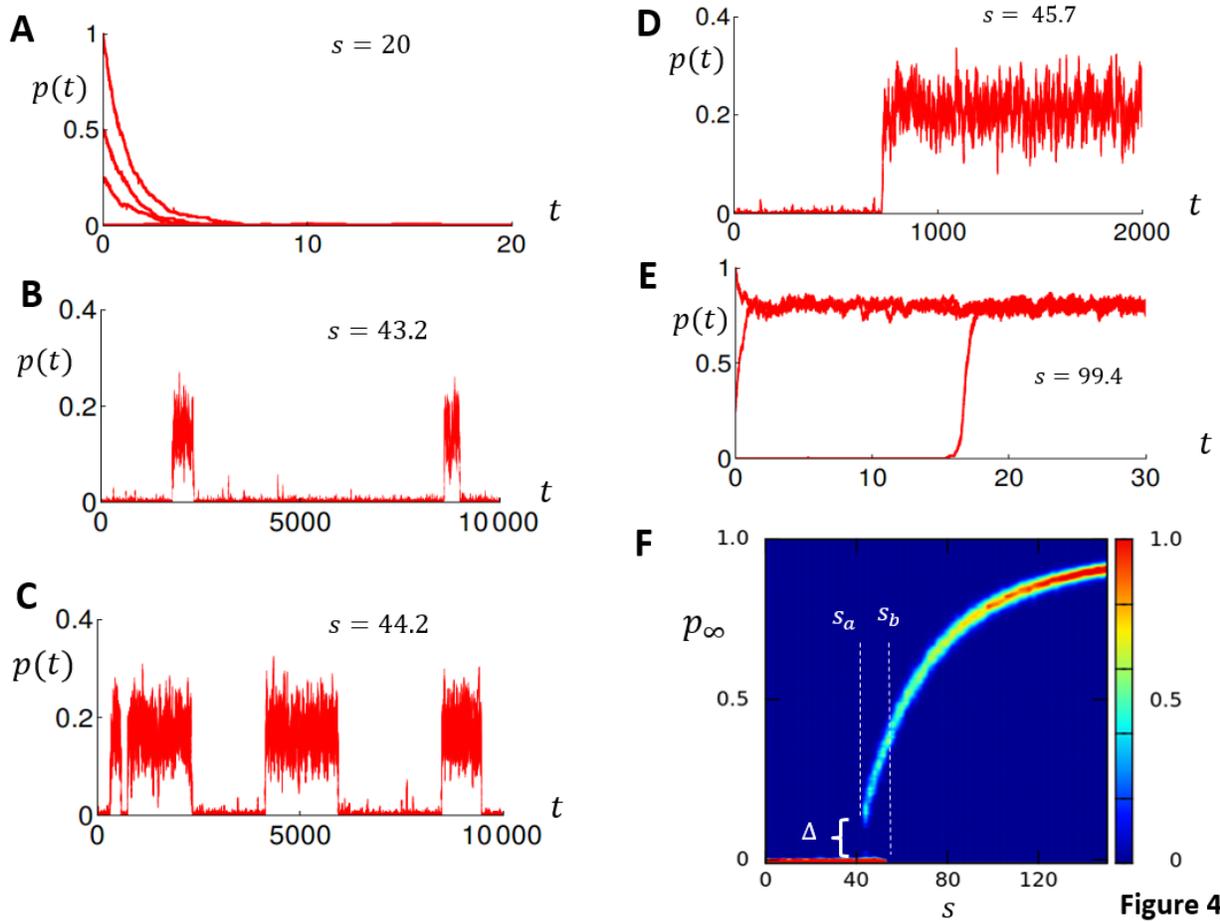

Figure 4

**Figure 5.** (A) Average waiting time to transition between inactive and active branch. Blue line denotes the waiting time $\langle t_w \rangle$ to transition from the inactive branch to the active branch. Black line denotes the average lifetime $\langle t_a \rangle$ spent on the active branch. The red line shows the averaged steady state probability $\langle p_\infty \rangle$ corresponding to the active and inactive branch. (B) The gap at onset $\Delta$ vs network parameter $m$. (C) Density plot of the steady state probability $p_\infty$ of a BA network with $m = 2$. Network size is $N = 500$ with excitability $\eta = 0.001$.

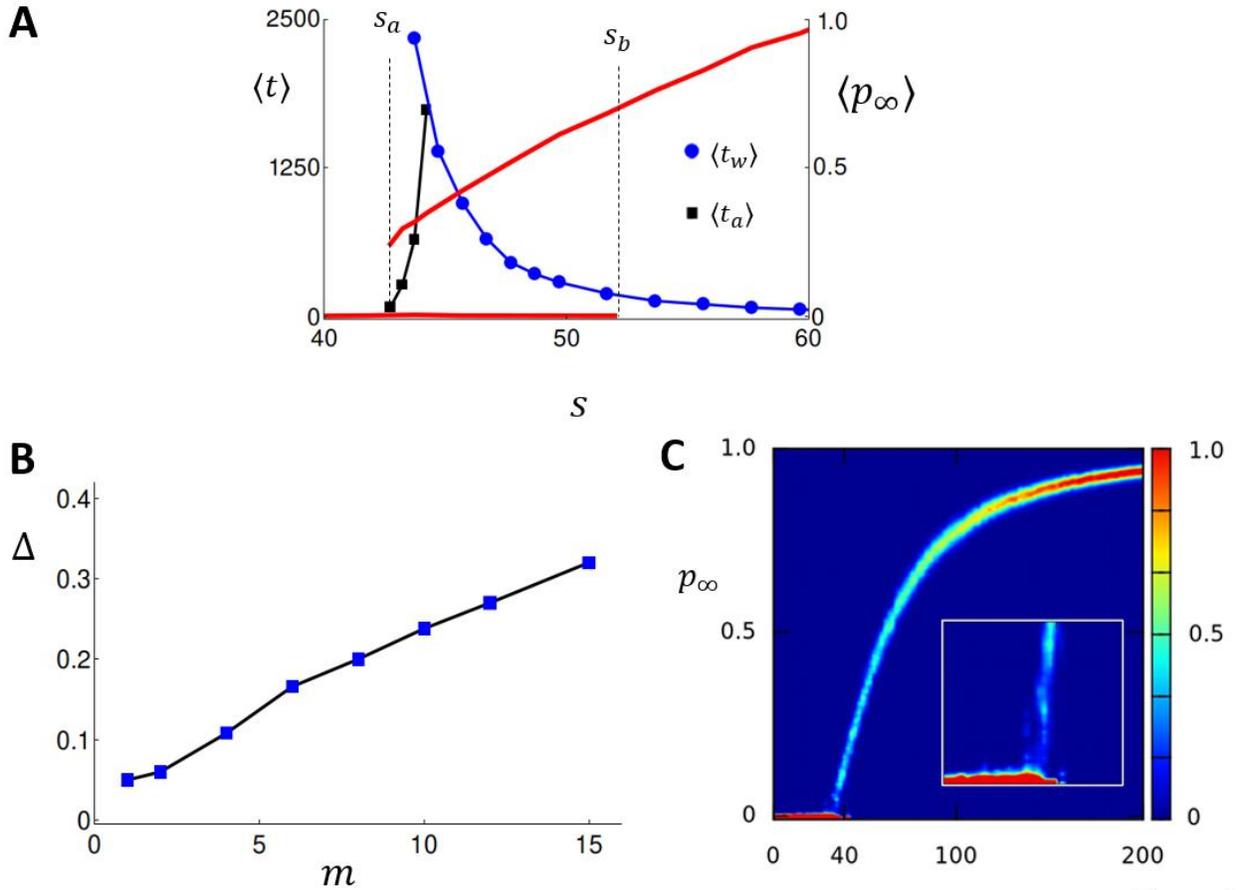

**Figure 5**

**Figure 6.** (A) Steady state solutions to the mean field equations with $\eta = 0.005$ and $\gamma = 2$. (B) Mean field approximation of steady state probability $p_\infty$ as a function of $s$ of an ER network above (red) and below percolation threshold (black). (C) Mean field approximation of steady state probability $p_\infty$ as a function of $s$ for a BA network with connectivity $m = 5$. In this simulation we use $\eta = 0.003$. (D) Plot of the gap $\Delta$ vs $m$.

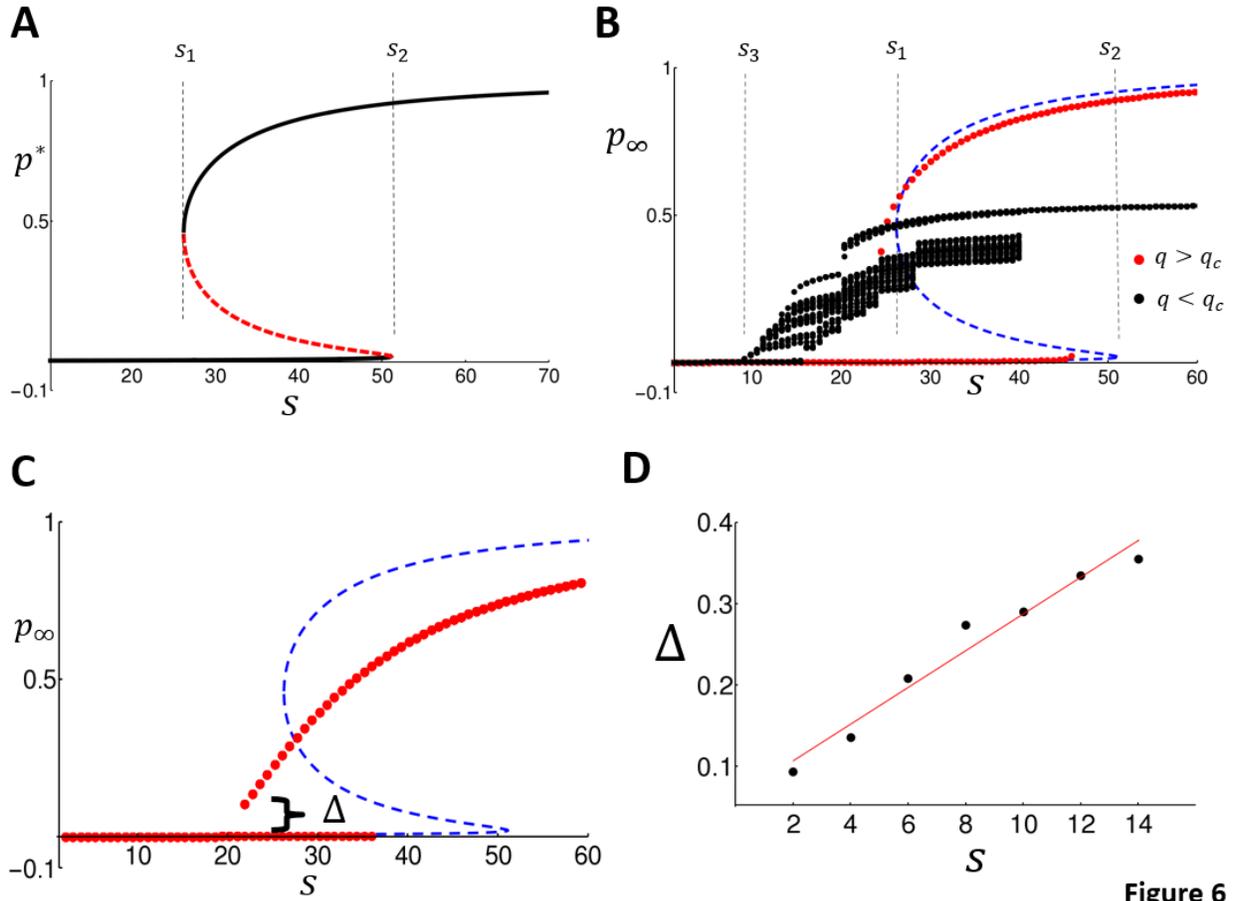

**Figure 7.** (A) Normalized ordered components of the leading eigenvector ($\phi_{1i}$), and activity level at node $i$ at steady state ($\bar{\eta}_i$), for a BA network with $m = 2$ and $s = 40$. Only the 50 largest components are shown, and all parameters are the same as in Figure 5C. (B) Network structure representing all nodes with $\phi_{1i} > 0.2$ in the center and the high activity sites with $\bar{\eta}_i > 0.5$ highlighted in red.

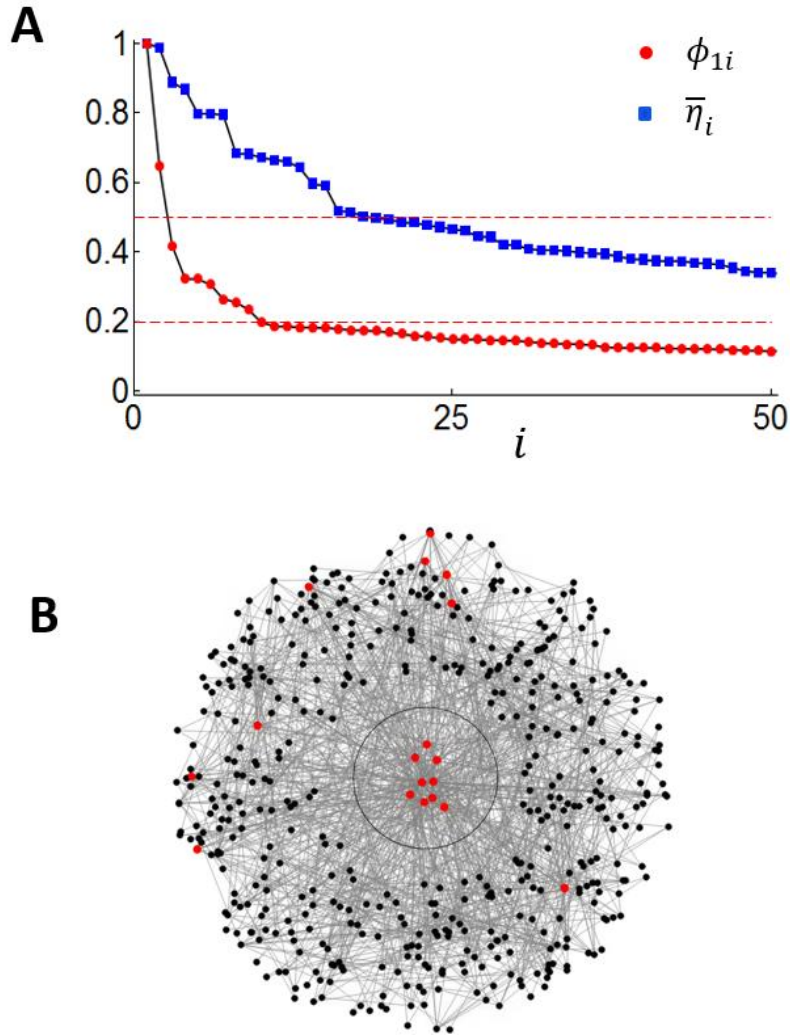

Figure 7

**Figure 8.** (A) Normalized ordered components of the leading eigenvector ($\phi_{1i}$), and activity level at node $i$ at steady state $\bar{\eta}_i$, for an ER network below the percolation threshold. Parameters are identical to that used in Figure 2D. (B) Network structure showing eigenvectors corresponding to the 3 largest eigenvalues (enclosed in dashed lines). High activity sites with $\bar{\eta}_i > 0.5$ are highlighted in red. Here, the network structure shows only the nodes that have more than two connections. The remaining weakly connected nodes were not active at steady state and are not shown.

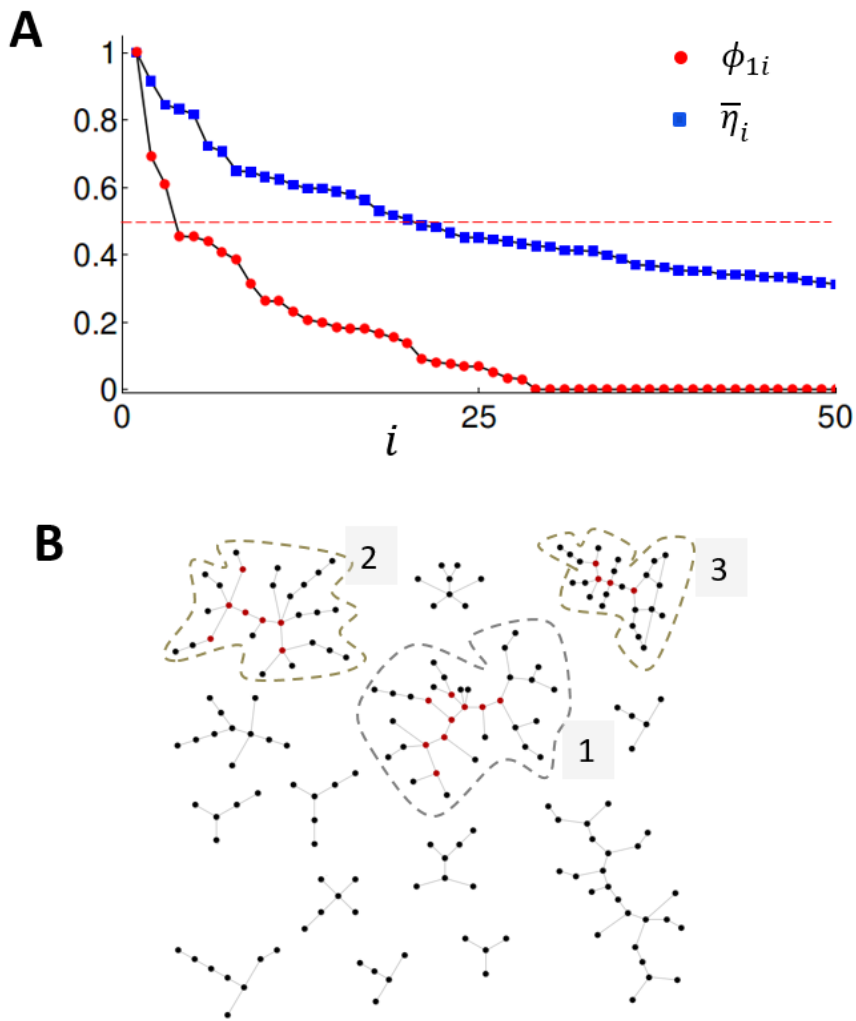

Figure 8

**Figure 9.** Localization of the leading eigenvector in the ER network. (A) Leading cluster size $\Gamma = n_1/N$ as a function of the linking probability $q$. $\Gamma$ is averaged over 80 independent network configurations. (B) Ordered components of the leading eigenvector above $(q > q_c)$ and below $(q < q_c)$ percolation threshold for a particular realization of an ER network with $N = 500$. (C) Leading cluster size $\Gamma = n_1/N$ as a function of system size $N$ below the percolation threshold $(q = 1/2N)$. (D) Leading cluster size $\Gamma = n_1/N$ as a function of system size $N$ above the percolation threshold $(q = 3/N)$. For C & D points are computed by averaging over 500 independent network configurations.

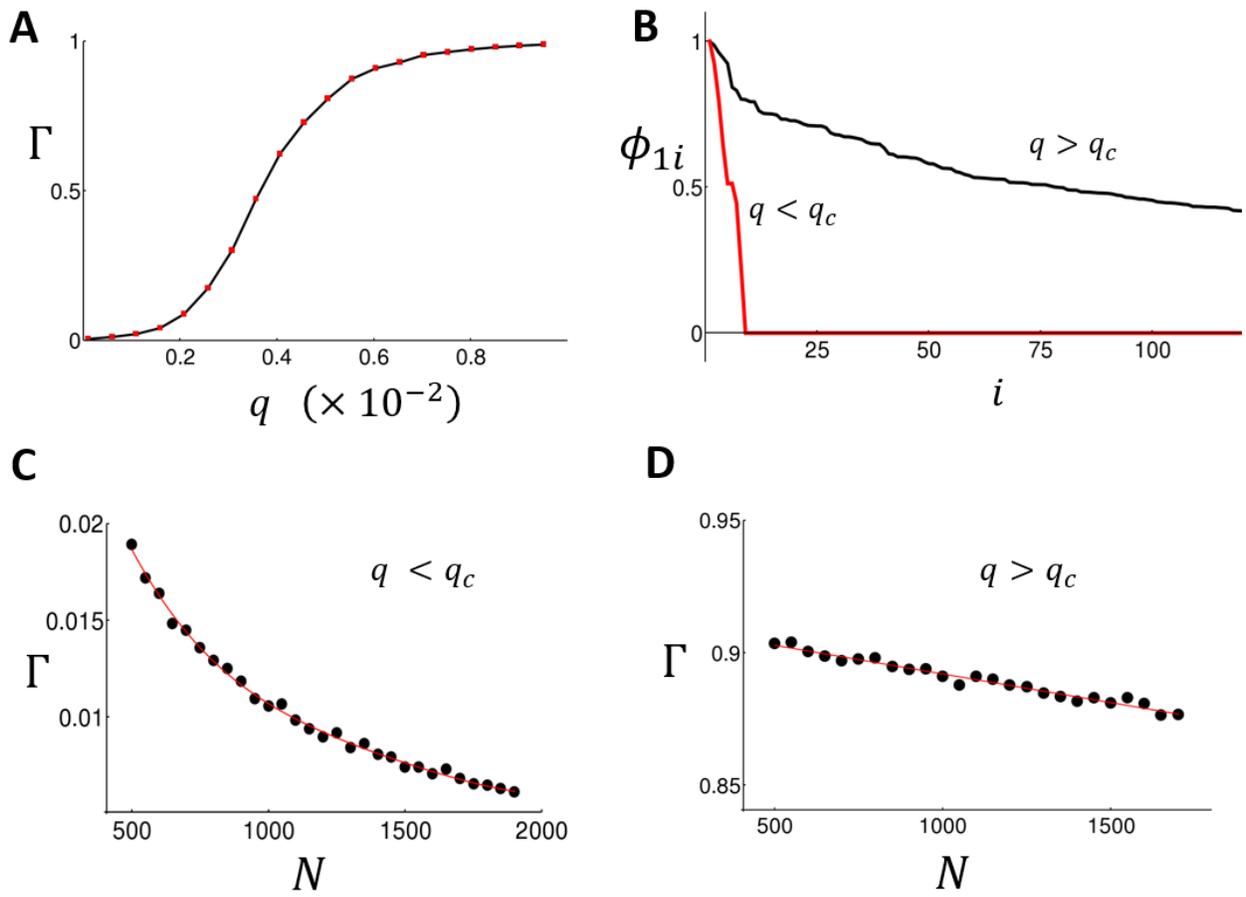

**Figure 9**

**Figure 10.** Structure of the leading eigenvector in BA networks. (A) Leading cluster size $\Gamma = n_1/N$ as a function connectivity $m$. (B) Leading cluster size $\Gamma = n_1/N$ as a function system size $N$. Results are averaged over 150 network configurations.

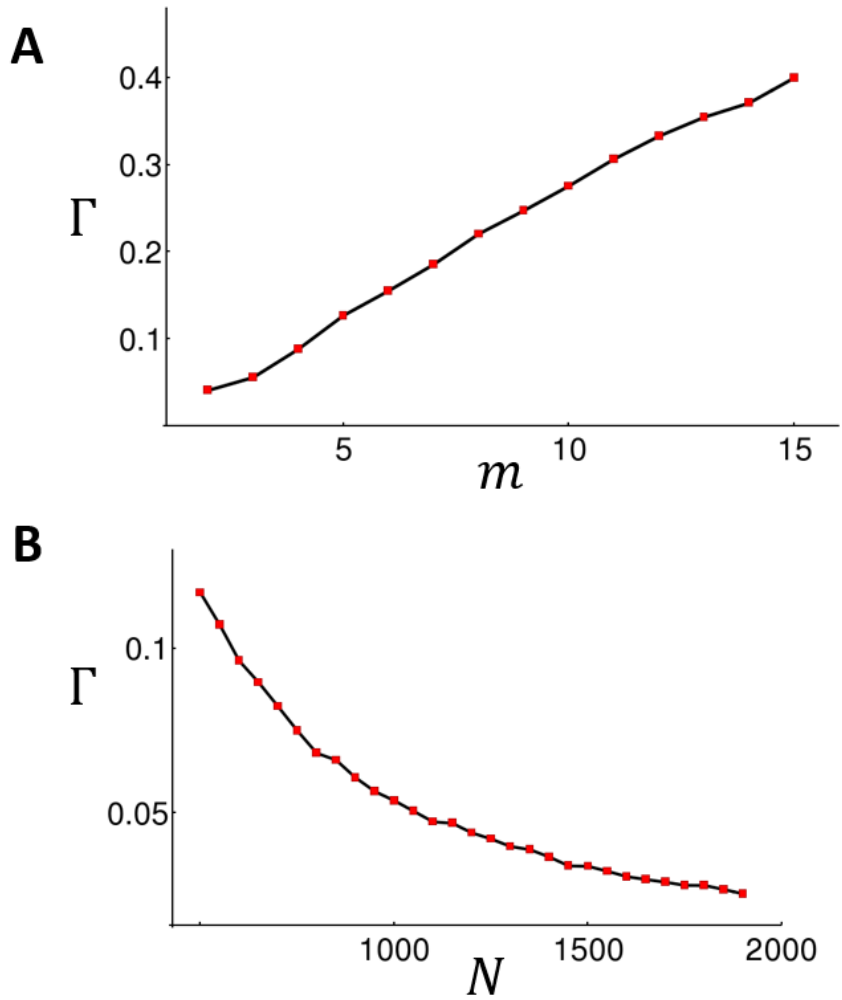

**Figure 10**

**Figure 11.** Signal spreading. (A) Plot of the number of active nodes $n(t)$ as a function of time in the case where the network is inactive at $t = 0$. Here, we use a BA network with $m = 2$, system size $N = 500$ and $\eta = 0.001$. Connectivity is fixed at $s = 39.8$. (B) Network graph with leading cluster placed in the interior. Here the highlighted nodes correspond to the ten largest $P_i$ computed using $M = 2000$ simulations. The threshold for activation is fixed at $n_c = 20$. (C) Same as above with $s = 159.2$. (D) Highlighted nodes correspond to the top 10 nodes active at a threshold of $n_c = 1$.

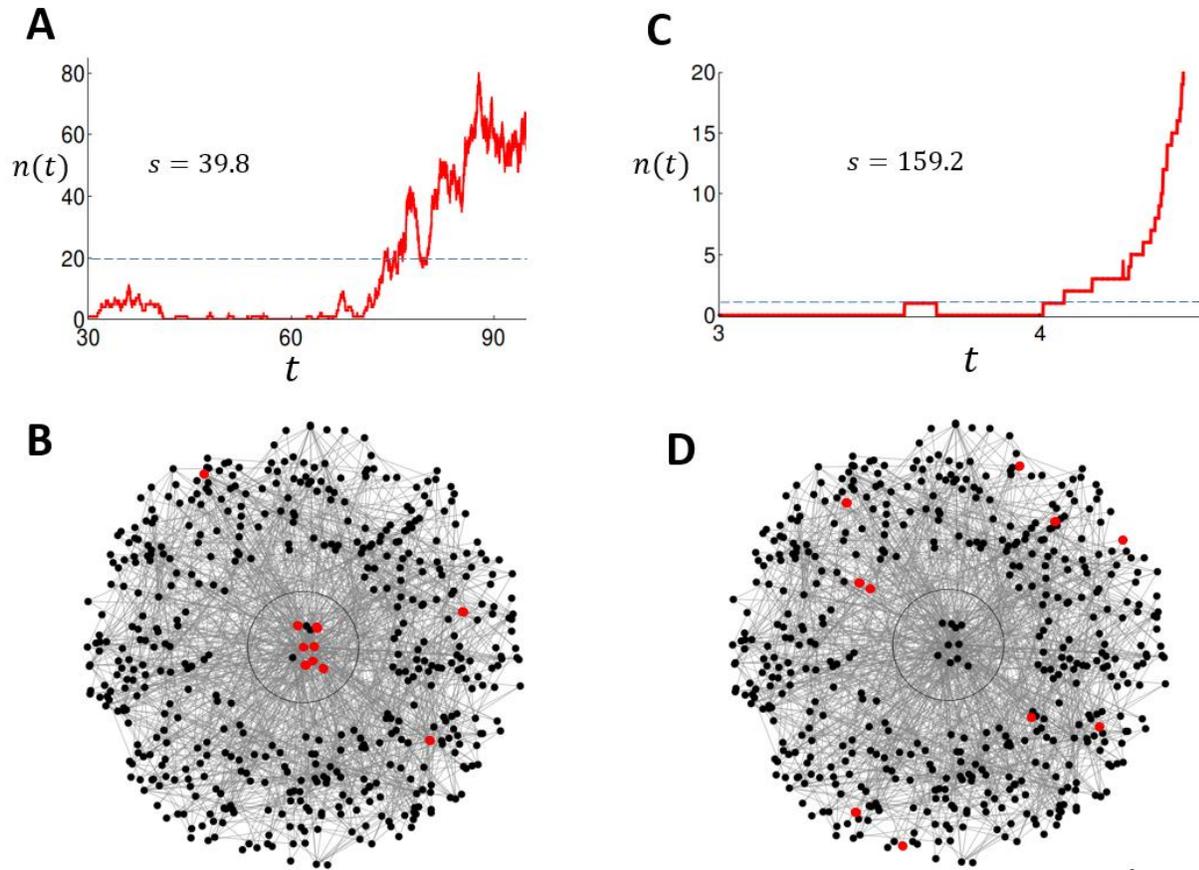

**Figure 11**

**Figure 12.** Pearson correlation between leading eigenvector and the distribution of active sites $\vec{P}$ at threshold as a function of the network connectivity. Points were computed using 2000 independent simulation runs. Red line indicates the average steady state $\langle p_\infty \rangle$ from the data shown in Figure 5C.

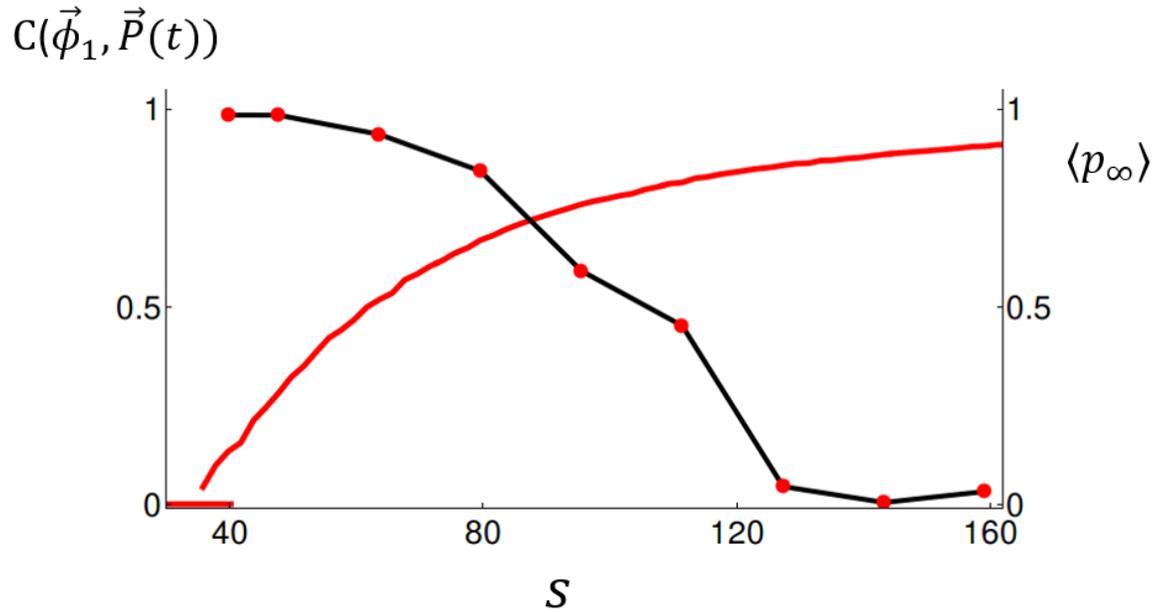

Figure 12